\RequirePackage{fix-cm}

\documentclass[twocolumn]{svjour3}         
\smartqed 
\usepackage{graphicx}

\usepackage{amsmath}
\usepackage{enumitem}
\usepackage{graphicx}
\usepackage{dcolumn}
\usepackage{array,multirow}
\usepackage{bm}
\usepackage{amsmath}
\usepackage{epstopdf}
\usepackage{amsfonts}
\usepackage{amssymb}
\usepackage{epsfig}
\usepackage{tabularx}
\usepackage{color}
\usepackage{xcolor}
\usepackage[colorlinks=true,citecolor=blue,urlcolor=cyan,breaklinks]{hyperref}
\usepackage{cite}

 \journalname{Draft}
\begin{document}

\title{Photons trajectories on a first order scale--dependent static BTZ black hole}


\author{Mohsen Fathi \and \'Angel Rinc\'on \and
        J.R. Villanueva 
}


\institute{Mohsen Fathi \at Instituto de F\'isica y Astronom\'ia, Universidad de Valpara\'iso, Avenida Gran Breta\~na 1111, Valpara\'iso, Chile\\
	 \email{mohsen.fathi@postgrado.uv.cl}
	\and \'Angel Rinc\'on \at
              Instituto de F{\'i}sica, Pontificia Universidad Cat{\'o}lica de Chile, Avenida Vicu{\~n}a Mackenna 4860, Santiago, Chile\\
              \email{arrincon@uc.cl}           
           \and
           J.R. Villanueva \at
               Instituto de F\'isica y Astronom\'ia, Universidad de Valpara\'iso, Avenida Gran Breta\~na 1111, Valpara\'iso, Chile\\
               \email{jose.villanueva@uv.cl}
}

\date{Received: date / Accepted: date}

\maketitle

\begin{abstract}
In this paper we study the motion of massless particles on a static BTZ black hole background in the context of scale--dependent gravity, which is characterized by the running parameter $\epsilon$. Thus, by using standard methods we obtain the equation of motions and then analytic solutions are found. The relevant non--trivial differences appear when we compare our solution against the classical counterpart.

\end{abstract}
\tableofcontents
\section{Introduction}

General Relativity (GR) predicts the existence of the so--called Black Holes (BHs). These objects play a dominant role in physics because we know they link not just gravity but also include quantum and statistical mechanics.
The Hawking's work \cite{Hawking:1974rv,Hawking:1974sw} showed that black holes indeed emit radiation from their horizon. This is one of the most important features which reveals that this objects are perfect to get insight in several directions in gravitational theories.
	In addition, BHs are parametrized by just a few constants: i) the mass, ii) the angular momentum and iii) the charge. Despite the conceptually interesting features that black hole exibit, was just after the LIGO direct detections of gravitational waves \cite{Abbott:2016blz,Abbott:2016nmj,Abbott:2017vtc} when them received considerable attention. Currently, one of the most relevant topics is concerning about the so--called Quasinormal modes (QNM) of black holes, which contains invaluable information regarding the aforementioned parameters of BH solution. To become familiar with, can see the classical reviews \cite{Kokkotas:1999bd,Berti:2009kk} and for more recent solutions see \cite{Panotopoulos:2018can,Panotopoulos:2017hns,Destounis:2018utr,Manfredi:2017xcv,Rincon:2018sgd,Rincon:2018ktz}.
	Given that black holes combine classical and quantum effects, the research of this kind of objects might help us to improve our understanding of how gravity and quantum mechanics work together.  
In particular, jut after the seminal work of Deser, Jackiw, 't Hooft and Witten 
\cite{Deser:1983tn,Deser:1988qn,tHooft:1988qqn,Witten:1988hc}, 
gravitation in (2+1) dimensions was considered as an ideal scenario to investigate conceptual issues such as the nature of observables and the ``problem of time" \cite{Carlip:1995qv}.
Thus, in order to check effects beyond classical Einstein gravity, the BTZ black hole (the first BH obtained with negative cosmological constant in that dimension) serve as a toy model to try to understand quantum gravity.
Originally, general relativity in (2+1) dimensions was not considered seriously. One of the main reasons is that it does not have a Newtonian limit \cite{0264-9381-3-4-010}, however, the pioneer BTZ solution showed that it is indeed a black hole and it is interesting to learn about it due: i) it has an event horizon ii) it appears as the
final state of collapsing matter, and finally iii) it has thermodynamic properties quiet similar to a (3+1)-dimensional black hole \cite{Carlip:1995qv}.

Thus, after the discovery of BTZ black hole solution \cite{Banados:1992wn,Banados:1992gq}
the idea of gravity in (2+1) dimension gained a lot of adepts
analizing several interesting properties usually treated in the (3+1) dimensional counterpart, for example: its geodesic structure \cite{Cruz:1994ir}, thermodynamic properties \cite{Carlip:1995qv,Banados:1998gg,Cruz:2004vp},  quasinormal modes \cite{Cardoso:2001hn,Crisostomo:2004hj,Panotopoulos:2018can}, stable and regular interior solutions that matches with a BTZ background \cite{Cruz:1994ar,garcia03,Cruz:2004tz,Cataldo:2006yk} scale--dependent solutions \cite{Koch:2016uso,Rincon:2018dsq}, among others.
In light of this, in the present paper, we will investigate the gravitational effects on light produced in the spacetime of the scale--dependent version of the classical BTZ black hole (recently, Rinc\'on \& Villanueva  have studied the Sagnac effect in these scenarios \cite{sdbtzsagnac}). The importance of this study is twofold: first, because the motion of light provides a way to classify an arbitrary spacetime (in order to reveal its structure) and second because the quantum features of this scale--dependent black hole could modify the classical trajectories of light. 

This article is organized as follow: after this brief introduction, we will discuss the fundamental ingredients of scale--dependent theory of gravity in Sect. \ref{SDT} and, after that, in Sect. \ref{Model} we will discuss briefly the scale--dependent solution in (2+1) dimensions. 
Then, in Sect. \ref{n_g} we obtain the equations of motion for massless particles by making use of the standard Lagrangian procedure on this static scale--dependent BTZ black hole solution, and finally in the last section we will briefly summarize the main result of this paper.

\section{Scale--Dependent Theory} \label{SDT}
	The so--called scale--dependent scenario has received considerable attention 
	in the context of black holes, wormholes, quasinormal modes as well as other applications. Given 
	that the philosophy beyond this method is novel, we will briefly summarize the main points regarding
	how this idea is applied.
	First, for a more detailed discussion see \cite{Koch:2016uso,Rincon:2017ypd,Rincon:2017goj,sdbtzsagnac,Rincon:2017ayr,Contreras:2017eza,Rincon:2018sgd,Hernandez-Arboleda:2018qdo,Contreras:2018dhs,Rincon:2018lyd,Rincon:2018dsq,Contreras:2018gct,Canales:2018tbn,Rincon:2019cix,Contreras:2018swc,Contreras:2018gpl}.

	The crucial point is that in the scale--dependent scenario, the couplings of a certain theory are not constant any more. Inspired by the asymptotic safety program we allow that the coupling evolve with certain energy scale. This assumption allows to extend the classical well--defined solutions to include quantum corrections which, by definition, are taken to be small. In our particular problem, we only have two coupling:
	%
	i) the 	Newton’s coupling $G_k$ and ii) the cosmological coupling $\Lambda_k$. 
	Please, be aware by noting that the Newton's coupling is related with the gravitational coupling using the simple relation $\kappa_k \equiv 8 \pi G_k$. 
	The problem have two independent fields: i) the arbitrary energy scale $k$ and ii) the metric field $g_{\mu \nu}$
	
	The effective action is then written as
	\begin{align}
	\Gamma[g_{\mu \nu}, k] \equiv \int \mathrm{d}^3 x \sqrt{-g}
	\Bigg[ 
	\frac{1}{2 \kappa_k} \Bigl(R - 2 \Lambda_k \Bigl) \ + \ \mathcal{L}_M
	\Bigg]   ,
	\end{align}
	where $\mathcal{L}_M$ is the Lagrangian density of the matter fields, and after varying the effective action with respect to the metric field, the effective Einstein field equations are as follows:
	
	\begin{align}
	G_{\mu \nu } + \Lambda_k g_{\mu \nu} \equiv \kappa_k T_{\mu \nu}^{\text{effec}}
	\end{align}
	where the effective energy momentum tensor is defined according to
	\begin{align}
	\kappa_k T_{\mu \nu}^{\text{effec}} &=  \kappa_k T_{\mu \nu}^{M} - \Delta t_{\mu \nu}.
	\end{align}
	The object $T_{\mu \nu}^{\text{effec}}$ now include two contributions, i.e. in addition to the usual matter content, we now have the non--matter source provided by the running of the gravitational coupling.
	This new tensor is then defined as:
	\begin{align}
	\Delta t_{\mu \nu} \equiv G_k \Bigl( g_{\mu \nu} \square - \nabla_{\mu} \nabla_{\nu} \Bigl) G_k^{-1}. 
	\end{align}
	Even though matter source is always an interesting ingredient in gravitational theories, we will focus on the simplest case in which $T_{\mu \nu}^{M} = 0$ to investigate the effect of the scale--dependent couplings into a the well--known BTZ black hole solution. Please, note that in some circumstances, the cosmological coupling is taken as a source term giving rise to $T_{\mu \nu}^{M} \neq 0$. This, however, is just a reinterpretation of the cosmological constant and does not provide a real source.
	
	The additional field $k(x)$ gives us an auxiliary equation to complete the set. Thus, the relation is obtained from the condition
	\begin{align}\label{eomk}
	    \frac{\delta \Gamma[g_{\mu \nu},k]}{\delta k} = 0
	\end{align}
	
	This restriction can be seen as an a posteriori condition towards background independence \cite{Stevenson:1981vj,Reuter:2003ca,Becker:2014qya,Dietz:2015owa,Labus:2016lkh,Morris:2016spn,Ohta:2017dsq}
	The Rel. \eqref{eomk} provides us a restriction between $G_k$ and $\Lambda_k$ which reveals that the cosmological parameter needs to be considered in order to obtain self--consistent scale--dependent solutions. Notice that if we consider an additional contribution
	i.e., if $\mathcal{L}_M \neq 0$, then the cosmological coupling is not mandatory. As we commented before, the above equation closes the system, but the implementation	of this is a difficult task.  
	
	To get physical information out of those equations one has to set the renormalization scale $k(x)$ in terms of the physical variables of the system under consideration $k\rightarrow k(x, \dots)$. This choise, however, breaks the reparametrization symmetry. In order to recover the aforementioned symmetry, and to circumvent the use of \eqref{eomk}, we can supplement the field equations by assuming some energy constraint. Usually we have four standard energy conditions which play a important role in GR. 
	Despite many times some of these conditions can be violated, in general a well--defined model/solution maintain the validity of, at least, one of the energy conditions. In general, of the four energy conditions, the so--called null energy condition (NEC) is the least restrictive of them. We take advantage of the extreme NEC condition to get
	\begin{align}
	    T_{\mu \nu}^{\text{effec}} \ell^\mu \ell^\nu  = -\Delta t_{\mu\nu}\ell^\mu \ell^\nu  \overset{!}{=} 0,
	\end{align}
	where $\ell^{\mu}$ is a null vector, taken similar to \cite{Rincon:2017ayr}. A clever choise of this vector allows us to get a differential equation for the gravitational coupling, namely
	\begin{align} \label{EDO_G}
	    G(r)\frac{\mathrm{d}^2G(r)}{\mathrm{d}r^2} - 2 \left(\frac{\mathrm{d}G(r)}{\mathrm{d}r}\right)^2 = 0
	\end{align}
	Solving the above differential equation we decrease a degree of freedom of the problem. After replacing $G(r)$ into the effective Einstein field equations we are able to obtain the functions involved.
	It is remarkable that, for the case of coordinate transformations we have
	\begin{align}
	\nabla^\mu G_{\mu \nu}=0.
	\end{align}
	In the next section we will briefly discuss a new black hole solution in the context of scale--dependent couplings inspired by quantum gravity reported in \cite{Koch:2016uso,Rincon:2018lyd}.

\section{The background: Static circularly symmetric black hole solutions} \label{Model}
The metric, in the absence of charge, adopts circular symmetry whereas the functions involved only have radial dependence. With this in mind, the line element defined in terms of the usual Schwarzschild coordinates ($ct, r, \phi$) is as follow
\begin{align}\label{lineelans}
\mathrm{d}s^2 &= -f(r)\, \mathrm{d} (c t)^2 + f(r)^{-1} \, \mathrm{d}r^2 + r^2 \mathrm{d}\phi^2.
\end{align}
where we need to found the metric funtion $f(r)$ and the cosmological coupling $\Lambda(r)$. Solving first \eqref{EDO_G} and then $\{f(r), \Lambda(r)\}$ we obtain

\begin{align} 
G(r) = \ \ &\frac{G_0}{1 + \epsilon r}
\label{G}
\\
f(r) = \ \ &-\frac{8 G_0 M_0}{c^2} Y(r) + \frac{r^2}{\ell_0^2},
\label{f}
\\
\begin{split}
\Lambda(r) = \ \ &-\frac{1}{\ell_0^2}\left(\frac{1+3\,\epsilon\, r}{1+\epsilon r}\right) + \frac{8M_0 G(r)}{c^2 r^2} Y(r) \ \times
   \\
   &
   \left[r \epsilon+\frac{1}{2}(1+2 r \epsilon) \left(\frac{\textrm{d}\ln Y(r)}{\textrm{d}\ln r}\right)\right].
    \label{lambda}
\end{split}
\end{align}
where $Y(r)$ is an auxiliary function defined as follow
\begin{equation}\label{Yder}
Y(r) \equiv 1 - 2 x + 2 x^2 \ln \bigg(1 + \frac{1}{x}\bigg),\quad x\equiv r \epsilon.
\end{equation}
The set of constants $(\cdots)_0$ are defined as the classical values, and $\epsilon$ is the parameter which encodes the scale--dependent corrections. We then have four integration constants: i) the gravitational coupling $G_0$, ii) the cosmological coupling $\Lambda_0 \equiv - \ell_0^{-2}$, iii) the classical mass $M_0$ and finally iv) the running parameter $\epsilon$.  

On the other hand, the non--rotating classical solution \cite{Banados:1992wn,Banados:1992gq},
should be obtained when $\epsilon$ is turned off,
i.e. 
\begin{align}
\lim_{\epsilon \rightarrow 0} G(r) &= G_0,
\\
\lim_{\epsilon \rightarrow 0} f(r) &= f_0(r) \equiv -\frac{8 M_0 G_0}{c^2} + \frac{r^2}{\ell_0^2},
\\
\lim_{\epsilon \rightarrow 0} \Lambda(r) &= \Lambda_0.
\end{align}
According to the fact that the exact solution for the scale--dependent problem is complicated, and taking into account that quantum correction should be small, we can expand the full solution up to first order in $\epsilon r$ to obtain:
%
\begin{align}\label{g}
G(r)& \approx G_0 \,(1 - \epsilon r),
\\
\label{fap}
f(r) & \approx \frac{r^2}{\ell_0^2} +16 M \epsilon r-8 M,
\\
\label{lam}
\Lambda(r)& \approx \Lambda_0 (1+2r\epsilon),
\end{align} where $M\equiv M_0/m_p$ is the dimensionless mass, and $m_p$ is the Planck mass in the (2+1) gravity given by \cite{Cruz:2004vp}
\begin{equation}
    \label{mpla}m_p=\frac{c^2}{G_0}.
\end{equation}
The event horizon can be obtained demanding that $f(r)=0$. Thus, from Eq. (\ref{fap}) we have two solutions:
\begin{align} \label{rHori}
    r_{\pm} &= \pm \ R_0 
    \Bigl[
    \sqrt{1 + (\epsilon R_0)^2} 
    \mp 
    (\epsilon R_0)
    \Bigl],
\end{align}
where only the positive root has physical meaning. Furthermore, the parameter $R_0 \equiv \sqrt{8M}\ell_0$ is the classical horizon (i.e. the event horizon when $\epsilon$ goes to zero). Again, taking a Taylor series for small $\epsilon$ we observe how the classical horizon is corrected by taking into account quantum effects
\begin{align}
    r_{+} \approx R_0 
    \bigg[
    1 - \epsilon R_0 + \frac{1}{2}(\epsilon R_0)^2 + \mathcal{O}(\epsilon^3)
    \bigg].
\end{align}

Notice that an important relation is obtained from Eq. (\ref{rHori}) 
\begin{equation}\label{relaimp}
    8M = \frac{r_+^2}{\ell_0^2}+16 M \epsilon r_+,
\end{equation}
and also, the lapse function can be written as
\begin{eqnarray}
    f(r)
    &=&\frac{1}{\ell_0^2}(r-r_+)(r-r_-).
    \label{lapsfin}
\end{eqnarray}

\section{Null Geodesics}
\label{n_g}
In order to investigate motion of test particles on a static scale--dependent BTZ black hole background using the standard Lagrangian procedure \cite{Chandrasekhar:579245,cov04,Villanueva:2018kem}, we write the Lagrangian associated to the line element (\ref{lineelans}) which, for photons, reads
\begin{equation}
    \label{lagr}2\mathcal{L}=-f(r)\,c^2 \dot{t}^2 + \dot{r}^2 f(r)^{-1} + r^2\,\dot{\phi}^2=0,
\end{equation}where a dot means derivative with respect to an affine parameter $\tau$ along the geodesic. Since $\{t, \phi\}$ are cyclic coordinates, the corresponding conjugate momenta are conserved, so that 

\begin{equation}
\label{conserv}\Pi_{t}=-f(r) c^2 \dot{t}=- \mathcal{E},\quad \Pi_{\phi}=r^2 \dot{\phi}=L,
\end{equation}
where $\mathcal{E}$ is a constant which cannot be associated with the energy (per unit of mass) since the spacetime is not asymptotically flat, whereas $L$ is the magnitude of the angular momentum.
Replacing Eqs. (\ref{conserv}) into Eq. (\ref{lagr}) and defining $E\equiv \mathcal{E}/c^2$ we obtain 
\begin{equation}\label{rt}
\left(\frac{\mathrm{d}r}{\mathrm{d}\tau}\right)^2=E^2-V_{\textrm{eff}}(r),
\end{equation}where $V_{\textrm{eff}}(r)$ is the effective potential given by
\begin{equation}
\label{effpot}V_{\textrm{eff}}(r)=L^2 \frac{f(r)}{r^2}.
\end{equation}
Moreover, the potential presents a maximum at $r_m =\epsilon^{-1}$ and has a value 

\begin{align}
    V_{\textrm{eff}}(r) \bigg|_{r=r_{m}} &= \left(\frac{L}{\ell_0}\right)^2 \Bigl[ 1 + (\epsilon R_0)^2 \Bigl].
\end{align}

\begin{figure*}[ht]
\centering
\includegraphics[width=0.49\textwidth]{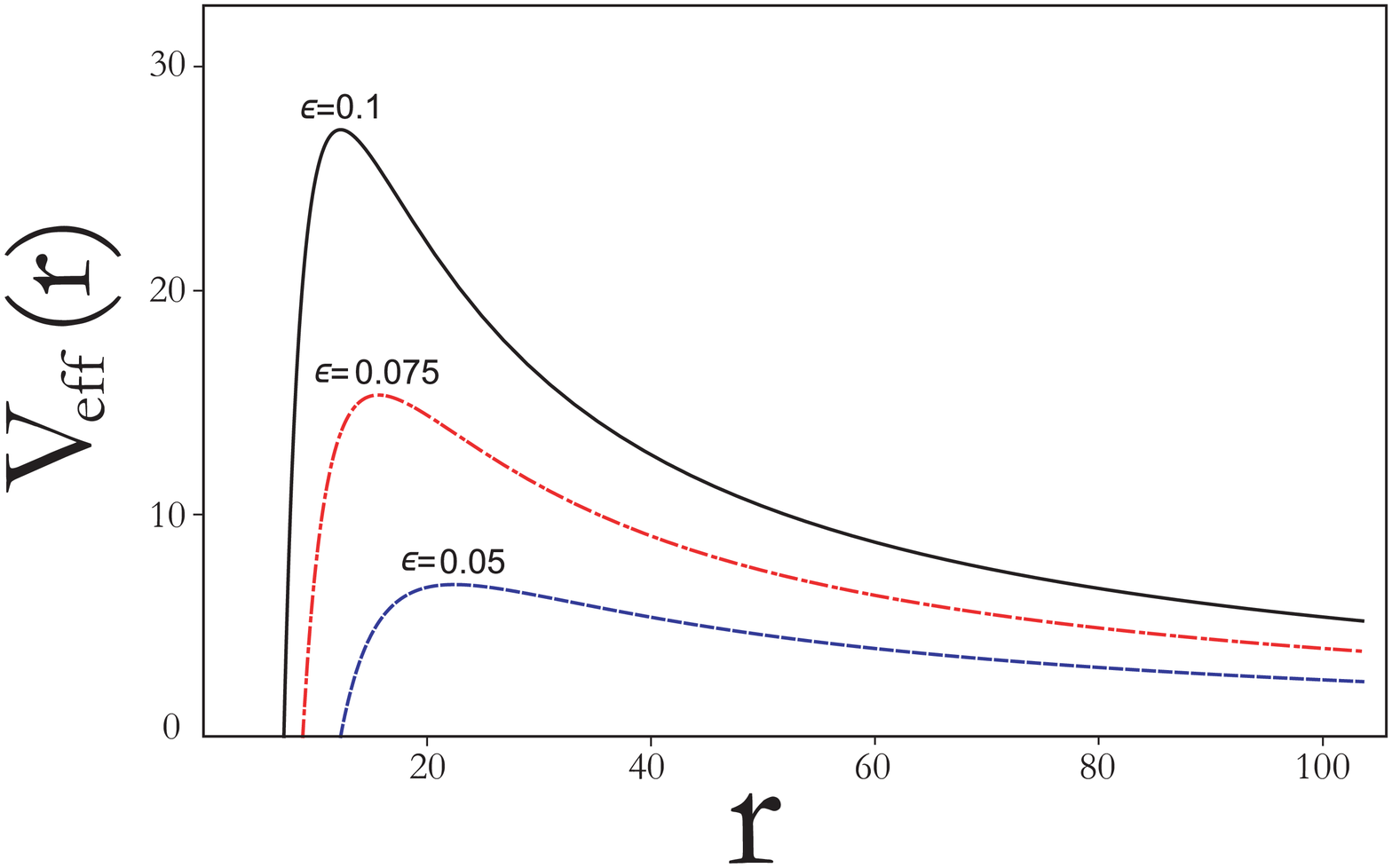}   \
\includegraphics[width=0.49\textwidth]{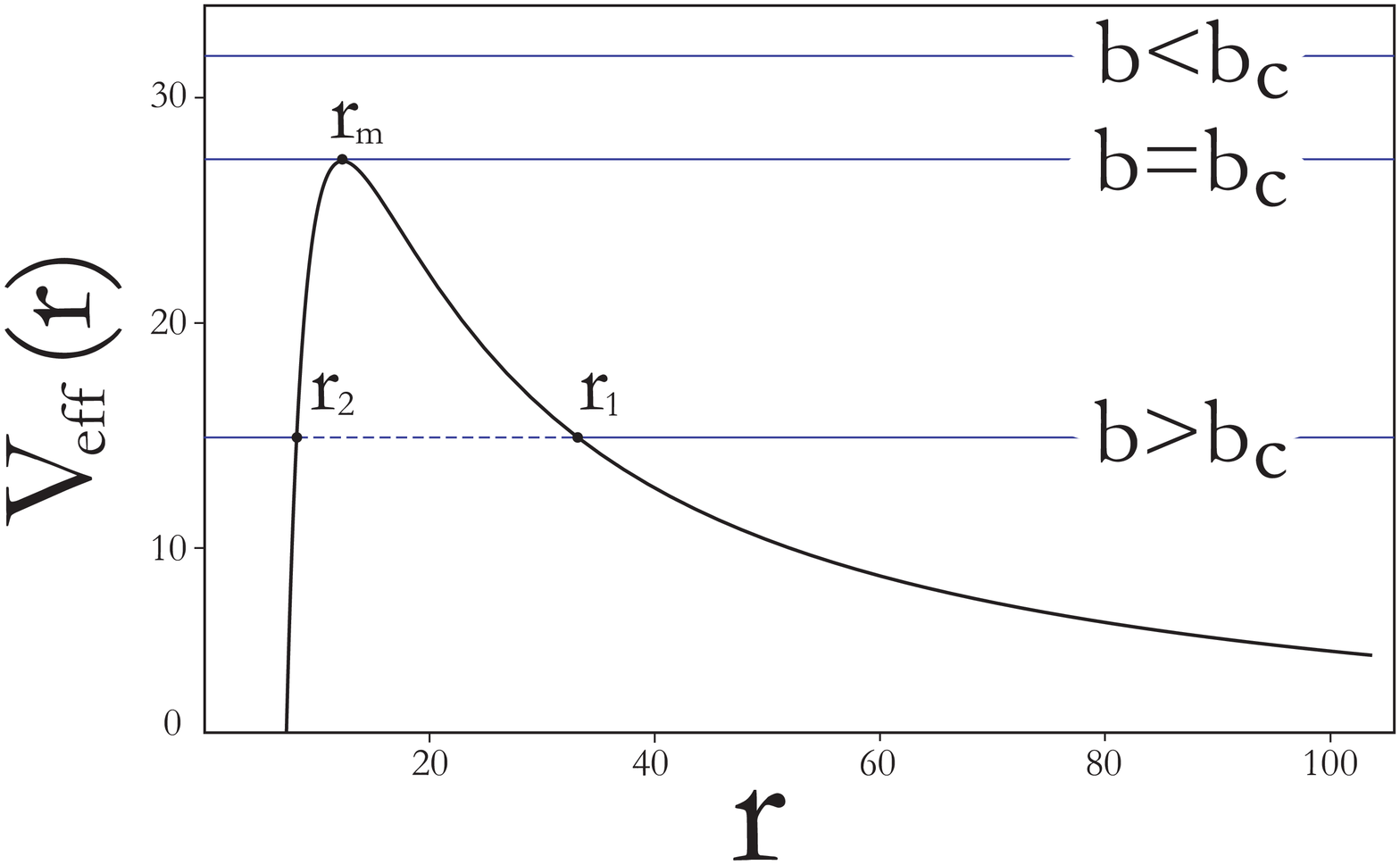}   \
	\caption{Plot for the effective potential  $V_{\textrm{eff}}$ as a function of the radial coordinate $r$, which presents a maximum equal to $V_{\textrm{eff}}=L^2/b_c^2$, where $b_c$ is the critical impact parameter given by Eq. (\ref{critimppar}), at $r_m=\epsilon^{-1}$. LEFT: Evolution of the effective potential for different values of the running parameter: $\epsilon=10^{-1}$, $\epsilon=7.5\times10^{-2}$ and $\epsilon=5\times 10^{-2}$, in arbitrary reciprocal length units. RIGHT: Depending on the value of the impact parameter, different trajectories are obtained. Thus, if $b>b_c$ there are two turning points, $r_1$ and $r_2$, which correspond to the periastron and apoastron distance for orbits of the first and second kind, respectively; geodesics with impact parameter $b=b_c$ allows an unstable circular orbit so photons arriving from infinity asymptotically approaches to the circle  of radius $r_m$ by spiralling. Also, from the opposite side, photons approaches to the same circle by spiralling around it; finally, if $b<b_c$ the motion is unbounded and photons coming from infinite goes to the event horizon and vice versa. }
	\label{figpotnull}
\end{figure*}

In Fig. \ref{figpotnull} this effective potential is plotted for different values of the parameters.
It is remarkable that the classical solution does not depend on the horizon radius, however, the quantum counterpart depends on the combination $\epsilon R_0$ which means that the maximum will be shifted when $\epsilon$ increases.
Using the second relation of Eq. (\ref{conserv}) and the chain rule, we obtain the radial--angular equation of motion, which is given by
\begin{equation}\label{rph}
L^2 \left(\frac{1}{r^2}\frac{\mathrm{d}r}{\mathrm{d}\phi}\right)^2=E^2-V_{\textrm{eff}}(r).
\end{equation}
Additionally,  Eq. (\ref{effpot}) together with Eqs. (\ref{rt}) and (\ref{rph}) allow us to obtain explicitly the equations of motion, which reads
\begin{equation}\label{rtn}
\left(\frac{\mathrm{d}r}{\mathrm{d}\tau}\right)^2
=
\left[E^2 - \left(\frac{L}{\ell_0}\right)^2 \right]-\frac{16 M \epsilon L^2}{r}+\frac{8 M L^2}{r^2},
\end{equation}and, with the change of variable $u=1/r$,
\begin{equation}\label{rphin}
\left(-\frac{\mathrm{d}u}{\mathrm{d}\phi}\right)^2=\left(\frac{1}{b^2}-\frac{1}{b_c^2}\right)+8 M (u-u_m)^2\equiv g(u),
\end{equation}where $u_m=\epsilon$, $b=L/E$ is the impact parameter and $b_c$ is a critical impact parameter, which corresponds to the value of the impact parameter for photons whose constant of motion $E^2=V_{\textrm{eff}}(r_{m})$, given by 

\begin{equation}
    \label{critimppar}
    b_c=\frac{\ell_0}{\sqrt{1 + (\epsilon R_0)^2}}.
\end{equation}
We observe that the critical value is, in this case, smaller than the classical counterpart.

\subsection{Radial Motion}\label{ng}

Photons with vanished angular momentum  $L=0$ have a zero effective potential, and then follow
a radial motion. By imposing this condition in Eq. (\ref{rtn}), it is straightforward to see that
\begin{equation}
\label{rtausl}r(\tau)=r_0\pm E \tau,
\end{equation}where $r_0$ is the location of the photon at $\tau=0$, and the plus (minus) sign indicates that the movement is made towards the spatial  infinity (event horizon). For the coordinate time, we use together Eqs. (\ref{conserv})-(\ref{rtn}) to obtain the following quadrature
\begin{equation}
    \label{radt}\frac{\mathrm{d}r}{\mathrm{d} t}=\frac{c}{\ell_0^2}(r-r_+)(r-r_-),
\end{equation}so an elementary integration  yields 
\begin{equation}
    \label{radtf}r(t)=\frac{r_+-\kappa_o r_-\,e^{\pm t/t_c}}{1-\kappa_o \,e^{\pm t/t_c}},
\end{equation}where 
\begin{equation}
    \label{rtfcon}\kappa_0=\frac{r_0-r_+}{r_0-r_-},\quad t_c=\frac{\ell_0^2}{c (r_+-r_-)}.
\end{equation}
\begin{figure}[ht]
\centering
\includegraphics[width=0.45\textwidth]{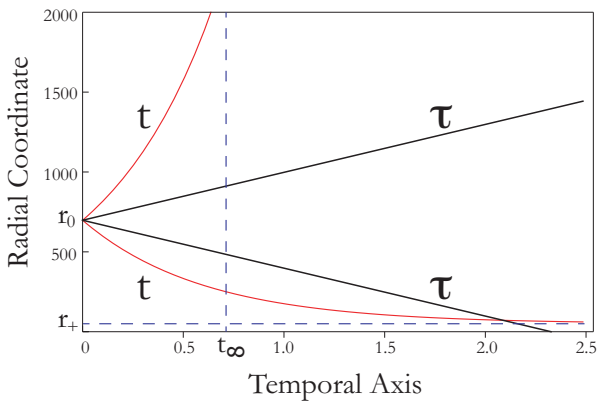} 
\caption{Plot for the radial coordinate as a function of the proper time $\tau$ and coordinate time $t$, described by Eqs. \eqref{rtausl} and \eqref{radtf}, respectively.}
\label{temp}
\end{figure}
Notice from Eq. (\ref{radtf}) and Fig. \ref{temp} that for an observer at infinity, photons take an infinite time to reach the horizon $r_+$ and a finite time 
\begin{align}
t_{\infty} &= t_c \,\ln \kappa_0^{-1}
\end{align}
to escape to infinity. This behaviour was reported before by Villanueva \& V\'asquez in the context of asymptotically Lifshitz spacetimes \cite{Villanueva:2013gra}. 
\subsection{The critical motion}
\begin{figure*}[ht]
\centering
\includegraphics[width=0.9\textwidth]{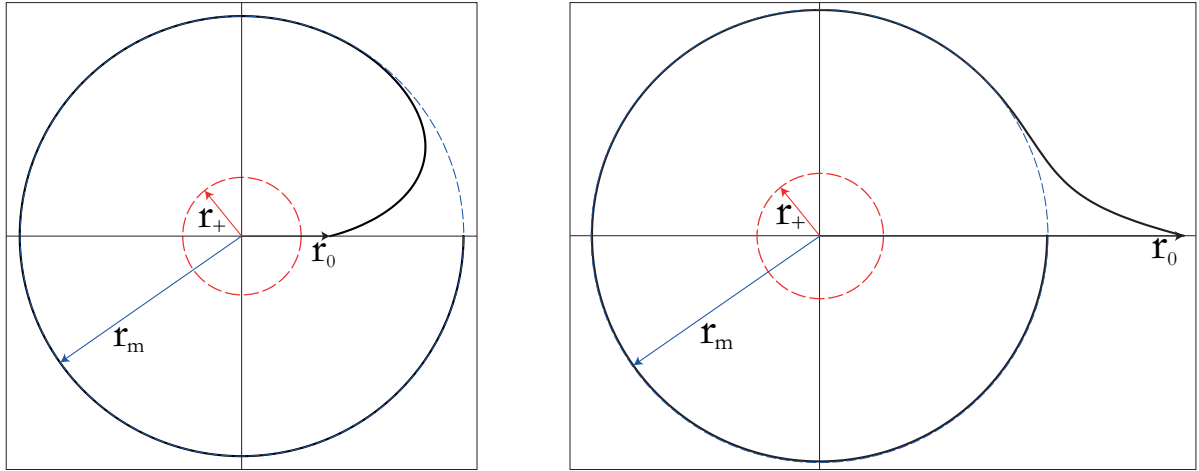} 
\caption{Polar plot for the critical motion of photons whose impact parameter is $b=b_c$. LEFT: Photons starting from a distance $r_0 < r_{m}$ approach asymptotically to the unstable circular orbit at $r_{m}$.
RIGHT: Photons starting from a distance $r_0>r_{m}$ approach asymptotically to the unstable circular orbit at $r_{m}$. Both graphs were made using $\epsilon=10^{-2}$ (in arbitrary reciprocal length units), $M=10$ and $\ell_0=100$ (in arbitrary length units), so that $r_m=100$, $b_c=11.11$ (in arbitrary length units).}
\label{crits}
\end{figure*}

Returning to the general equation (\ref{rphin}) we must distinguish the different possible cases based on the disposition of the roots of the polynomial $g(u)=0$. In order to obtain a qualitative analysis of the allowed motion we refers to the Fig.\ref{figpotnull}. Clearly in terms of the impact parameter $b$, there are three different allowed motion. The first one corresponds to the case $b=b_c$ so we have $g(u)=8M(u-u_m)^2=0$, so the motion corresponds to an asymptotic circular orbit (unstable) at $r=r_{m}$. Thus, considering 
$\phi=0$ when $r=r_0$, an integration of Eq. (\ref{rphin}) leads to
\begin{equation}
    \label{rcrit}
    r(\phi) = \frac{r_0}{\frac{r_0}{r_{m}}+\left(1-\frac{r_0}{r_{m}}\right)e^{\pm \sqrt{8M} \phi}},
\end{equation}
where the plus (minus) sign corresponds to the motion for which $r_0>r_{m}$ ($r_0<r_{m}$), and the corresponding polar plot are showed in Fig. \ref{crits}.
On the other hand, this kind of orbit allows to define a {\it cone of avoidance} whose generators are null rays \cite{Chandrasekhar:579245,cov04,Kuniyal:2015uta}. Denoting $\Psi$ as the half angle of the cone then,
\begin{equation}
    \label{cone1} \cot \Psi=\frac{1}{r}\frac{\textrm{d}\Tilde{r}}{\textrm{d}\phi}
\end{equation}where $\Tilde{r}$ is the proper length along the generators of the cone
\begin{equation}
    \label{cone2}\textrm{d}\Tilde{r}=\frac{\textrm{d}r}{\sqrt{f(r)}}=\frac{\ell_0\,\textrm{d}r}{\sqrt{(r-r_+)(r-r_-)}}.
\end{equation} Combining Eqs. (\ref{cone2}) and (\ref{rph}) with Eq. (\ref{cone1}) one obtains that
\begin{equation}
    \label{cone3} 
    \tan \Psi = 
    \left(\frac{r_{+}}{R_0} \frac{r_{-}}{R_0}\right)^{1/2} 
    \,\frac{\left(\frac{r}{r_+}-1\right)^{1/2}\,\left(\frac{r}{r_-}-1\right)^{1/2}}{\frac{r}{r_m}-1}.
\end{equation}
From this last equation it follows that
\begin{equation} \label{cone_Angel}
\Psi \rightarrow
\left\{
\begin{array}{lcl}
\sim \frac{r_m}{R_0} & \mbox{ \hspace{0.7cm} if \hspace{0.7cm} } & r \rightarrow  \infty 
\\ & & \\
= \frac{1}{2}\pi & \mbox{ if } & r = r_m
\\ & & \\
=0 & \mbox{ if } & r = r_{+}
\end{array}
\right.
\end{equation}

An important remark from the first of Eqs. \eqref{cone_Angel} is that the angle $\Psi$ goes to a constant value, which is radically different to the (3+1) gravity where $\Psi\sim 1/r$ as $r\rightarrow \infty$.

\subsection{The Bounded Motion}

\begin{figure*}[ht]
\centering
\includegraphics[width=0.95\textwidth]{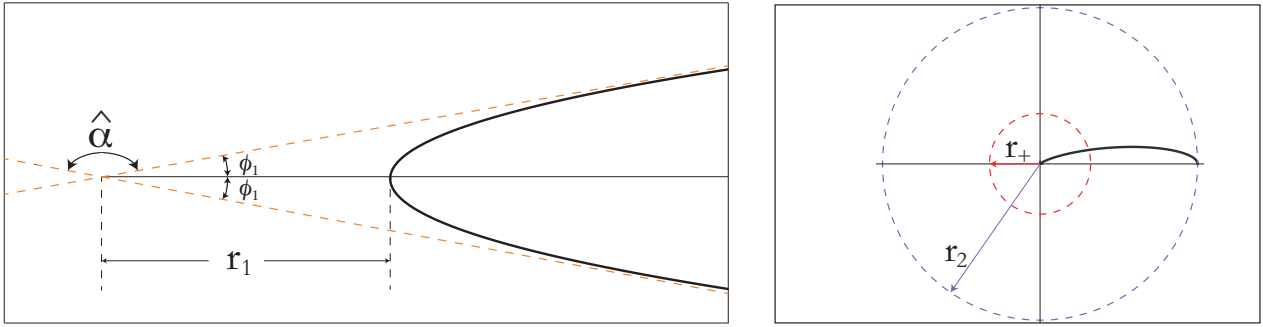}   \
\caption{Polar plot for the bounded motion of photons whose impact parameter is $b>b_c$. LEFT: Orbits of the first kind for photons whose radial coordinate is always greater than the periastron distance $r_1$, and the motion is symmetric with respect to $r_1$ so the  deflection angle result to be $\widehat{\alpha}=\pi-2\phi_1$. RIGHT: Orbits of the second kind for photons whose radial coordinate is always smaller than the apoastron distance $r_2$. Both graphs were made using $\epsilon=10^{-2}$ (in arbitrary reciprocal length units), $M=10$ and $\ell_0=100$ (in arbitrary length units), so that $r_m=100$, $b_c=11.11$ (in arbitrary length units).}
\label{first}
\end{figure*}

As is shows in Fig. \ref{figpotnull}, in the case when $b_c<b<\infty$ there are two kinds of allowed orbits:  the orbits of the first kind where $r_1<r$, so photons are scattered and reaches the spatial infinity, and orbits of the second kind for which the turning point satisfies the condition $r_2>r$, so photons cannot escape the capture zone and fall inexorably to the event horizon. The values of these turning points are obtained from the condition $E^2=V_{\textrm{eff}}$, and are given by
\begin{equation}
    \label{rot1}
    r_1=\frac{r_m}{1-\varepsilon},\quad r_2=\frac{r_m}{1+\varepsilon},
\end{equation}where $\varepsilon$ is the eccentricity given by
\begin{equation}
    \label{eccentr}\varepsilon=\frac{r_m}{\sqrt{8M}\mathcal{D}},
\end{equation}and $\mathcal{D}$ is the anomalous impact parameter given by the relation
\begin{equation}
    \label{anoimp}\frac{1}{\mathcal{D}^2}=\frac{1}{b_{c}^2}-\frac{1}{b^2}.
\end{equation}
Therefore, for orbits of the first kind with $\phi=0$ at $r=r_1$, a quick integration of Eq. (\ref{rphin}) yields 
\begin{equation}
    \label{pol1}
    r(\phi)=\frac{r_m}{1 - \varepsilon \cosh \left(\sqrt{8M} \phi\right)},
\end{equation} 
which is depicted in Fig.\eqref{first}.

\begin{figure*}[ht]
\centering
\includegraphics[width=0.48\textwidth]{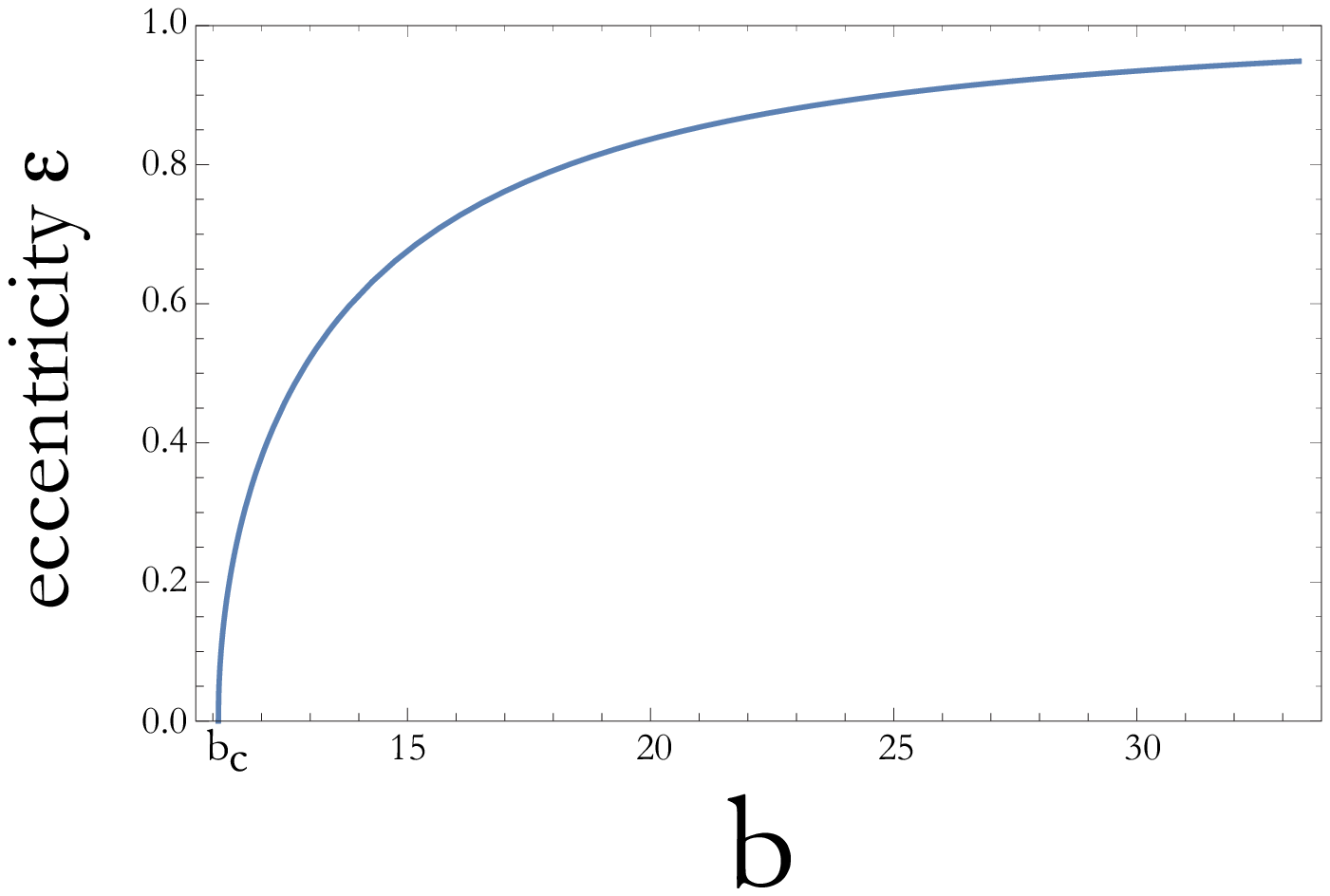}   \
\includegraphics[width=0.48\textwidth]{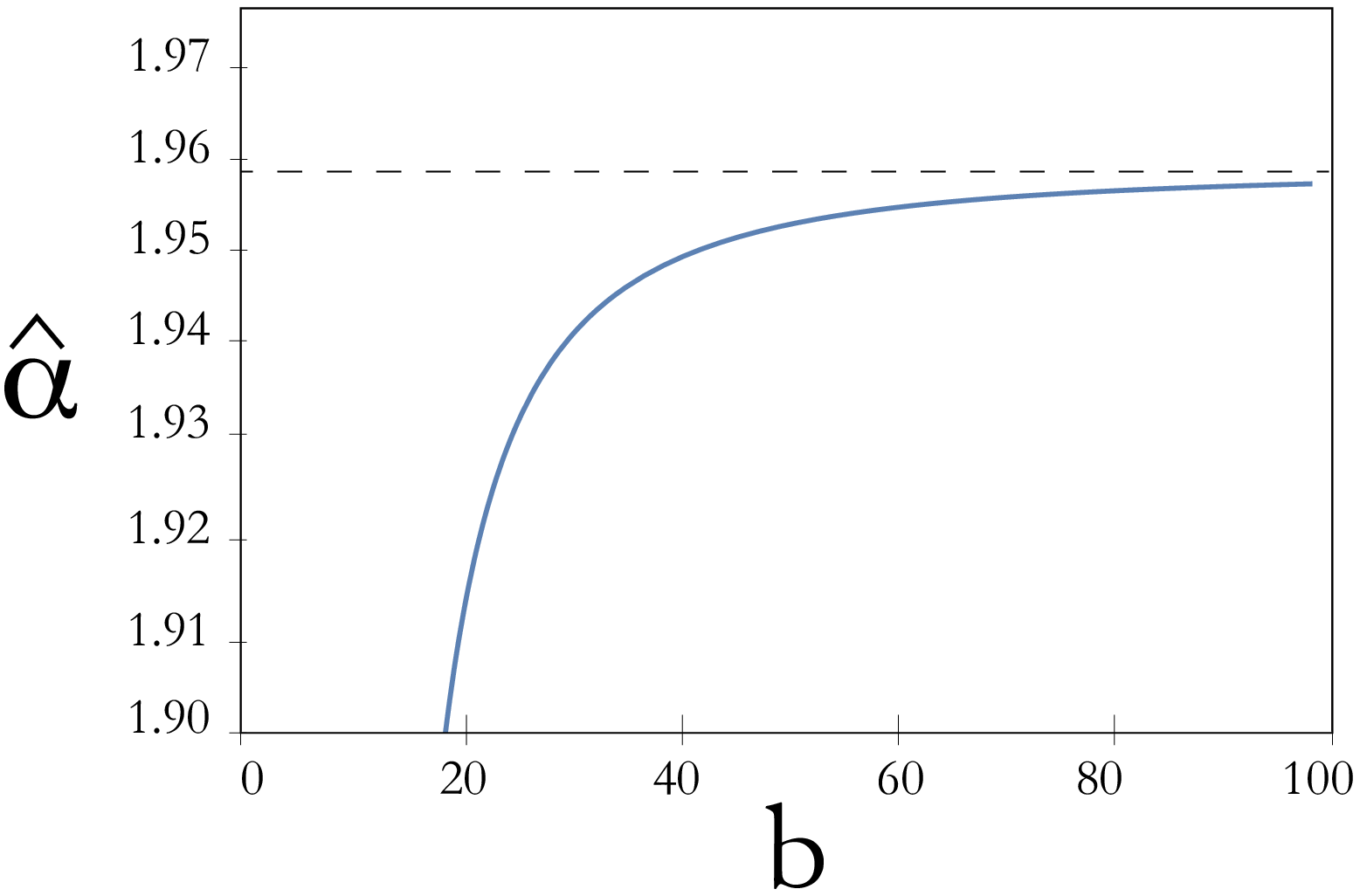}   \
\caption{LEFT: Eccentricity parameter $\varepsilon$ as a function of the impact parameter $b$. Notice that for the critical impact parameter corresponds the zero value of the eccentricity.
RIGHT: Deflection angle $\widehat{\alpha}$ as a function of the impact parameter $b$. Both graphs were made using $\epsilon=10^{-2}$ (in arbitrary reciprocal length units), $M=10$ and $\ell_0=100$ (in arbitrary length units), so that $r_m=100$, $b_c=11.11$ (in arbitrary length units).}
\label{deflecc}
\end{figure*}

Note that the test particles reach the infinity for an angle $\phi=\pm \phi_1$, given by
\begin{equation}
    \label{angperi}\phi_1=\frac{1}{\sqrt{8M}}\textrm{arccosh}\left(\frac{1}{\varepsilon}\right),
\end{equation} so the deflection angle $\widehat{\alpha}=\pi-2\phi_1$ becomes
\begin{equation}
    \label{defl}\widehat{\alpha}(b)=\pi-\frac{1}{\sqrt{2M}}\textrm{arccosh}\left(\frac{\sqrt{8M} b_c}{r_m} \frac{1}{\sqrt{1-(b_c/b)^2}}\right).
\end{equation}
In Fig. \ref{deflecc} the the eccentricity \eqref{eccentr} and deflection angle \eqref{defl} are plotted as a function of the impact parameter $b$.
 On the other hand, orbits of the second kind for which $\phi=0$ at $r=r_2$, are described for the following trajectory
 \begin{equation}
    \label{pol2}r(\phi)=\frac{r_m}{1 + \varepsilon \cosh \left(\sqrt{8M} \phi\right)},
\end{equation} which is depicted in the right panel of Fig. \ref{first} and, obviously, depends on the same parameter as orbits of the first kind.
\subsection{The Unbounded Motion}
\begin{figure*}[ht]
\centering
\includegraphics[width=0.95\textwidth]{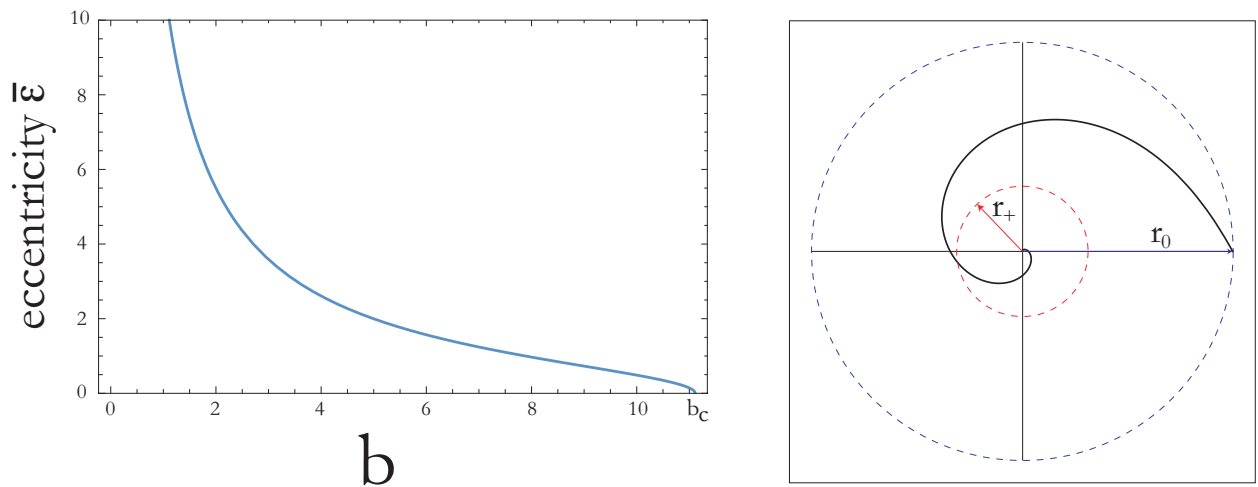}
\caption{LEFT: Eccentricity parameter $\Bar{\varepsilon}$ as a function of the impact parameter $b$ for unbounded motion. The critical impact parameter corresponds the zero value of the eccentricity and its tends to infinity as $b\rightarrow0$.
RIGHT: Trajectory follows by photons in an unbounded motion in which $\phi=0$ at $r=r_0$. Both graphs were made using $\epsilon=10^{-2}$ (in arbitrary reciprocal length units), $M=10$ and $\ell_0=100$ (in arbitrary length units), so that $r_m=100$, $b_c=11.11$ (in arbitrary length units).}
\label{unbfig}
\end{figure*}

When $0<b<b_c$ the polynomial $g(u)$ possesses a complex pair conjugate so that the motion is unbounded. This means that photons fall to the event horizon (or, depending on the initial conditions, to the spatial infinity) from a finite distance $r_0$. Therefore, for test particles coming from $r_0>r_m$ where $\phi=0$, the trajectory is described by 
\begin{equation}
    \label{trun}r(\phi)=\frac{r_m}{1+\Bar{\varepsilon}\, \sinh (\sqrt{8M}\phi-\varphi)}.
\end{equation}Here $\Bar{\varepsilon}$ is the eccentricity associated to the unbounded motion and is given by
\begin{equation}
    \label{eccunb}\Bar{\varepsilon}=\frac{r_m}{\sqrt{8M}\bar{\mathcal{D}}},\quad\textrm{with}\quad \frac{1}{\Bar{\mathcal{D}}^2}=\frac{1}{b^2}-\frac{1}{b_{c}^2},
\end{equation}and $\varphi$ depends on the initial position according to
\begin{equation}
    \label{varphunb}\varphi=\textrm{arcsinh}\left[\Bar{\varepsilon}^{-1}\left(1-\frac{r_m}{r_0}\right)\right].
\end{equation}Notice from Eq. (\ref{eccunb}) that now the range of the eccentricity is $0<\Bar{\varepsilon}<\infty$, as shown in the left panel of Fig. \ref{unbfig}. Also, in the right panel of the same graph the unbounded trajectory is plotted.

\section{Conclusions} \label{Conclusions}
In this contribution we have studied some relevant aspects of the geodesic structure of a scale--dependent BTZ black hole without angular momentum. Since the value of running parameter $\epsilon$ is small, we can justify the first--order work on this parameter so that we can analytically study the different possible trajectories followed by massless particles. 

The radial motion presents some similar feature as the standard black holes: for an observer at infinity, photons take an infinite (coordinate) time to reach the horizon even though in its proper system they cross the event horizon in a finite (proper) time. On the other hand, the observer will see that photons arrive at spatial infinity in a finite time $t_{\infty}=t_c\ln \kappa_0^{-1}$, where $t_c$ and $\kappa_0$ are given by Eq. \eqref{rtfcon}. This last feature was reported before by Villanueva \& V\'asquez \cite{Villanueva:2013gra} in the context of the Lifshitz space-times.

The angular motion is completely different to the standard non-rotating BTZ black hole. It is due to the existence of the extra term 16$M \epsilon r$ in Eq. (\ref{fap}).
Precisely, the scale--dependent BTZ black hole provides us of more complex physical situations which are absent in its classical counterpart. The linear term proportional to $\epsilon r$ plays a crucial role. Thus, given the structure of the lapse function, the effective potential has a maximum (located at $r_m = \epsilon^{-1}$, see Fig. \ref{figpotnull}) and therefore three well--defined regions, i.e. i) $b=b_c$ (critical trajectories), ii) $b>b_c$ (bounded trajectories), and finally iii) $b < b_c$ (unbounded trajectories).
In the critical motion, we found that photons  approaches the circle of radius $r_m$, asymptotically, by spiralling around it with an infinite number of times. Also, the cone of avoidance is calculated which is radically different to the (3+1) gravity where the half angle of the cone $\Psi\sim 1/r$ as $r\rightarrow \infty$, whereas in our case the angle goes to a non zero value $\Psi\sim r_m/R_0$ in that limit.
For the bounded motion we have calculated analytically the orbits of the first kind whose radial coordinate satisfies the condition $r>r_1>r_m$, where $r_1$ is the periastron distance given by Eq. (\ref{rot1}), and orbits of the second kind for which the relation $r<r_2<r_m$ is satisfied, where $r_2$ is the apoastron distance. Both orbits strongly depend on the anomalous impact parameter $\mathcal{D}$ (c.f. Eq. (\ref{anoimp})), across the eccentricity parameter $0<\varepsilon<1$, given by Eq. (\ref{eccentr}) (c.f. Fig. \ref{first}). Also, the deflection angle $\widehat{\alpha}$ for orbits of the first kind is analytically calculated in terms of the impact parameter $b$. Both quantities, $\varepsilon$ and $\widehat{\alpha}$, are plotted as a function of $b$ in Fig. \ref{deflecc}.
Finally, the unbounded motion is studied analytically resulting in the trajectories defined by Eq. \eqref{trun}, which depend on the impact parameter via the eccentricity $0<\bar{\varepsilon}<\infty$ given by Eq. \eqref{eccunb}, see Fig. \ref{unbfig}.

\begin{acknowledgements}
The work of A.R. was supported by the CONICYT-PCHA/Doctorado Nacional/2015-21151658 and the work of J.V. was partially supported by the Centro de Astrof\'isica de Valpara\'iso (CAV).\end{acknowledgements}

\bibliographystyle{spphys} 
\bibliography{Biblio_v1.bib}

\begin{thebibliography}{10}
\providecommand{\url}[1]{{#1}}
\providecommand{\urlprefix}{URL }
\expandafter\ifx\csname urlstyle\endcsname\relax
  \providecommand{\doi}[1]{DOI \discretionary{}{}{}#1}\else
  \providecommand{\doi}{DOI \discretionary{}{}{}\begingroup
  \urlstyle{rm}\Url}\fi

\bibitem{Hawking:1974rv}
S.W. Hawking, Nature \textbf{248}, 30 (1974).
\newblock \doi{10.1038/248030a0}

\bibitem{Hawking:1974sw}
S.W. Hawking, Commun. Math. Phys. \textbf{43}, 199 (1975).
\newblock \doi{10.1007/BF02345020, 10.1007/BF01608497}.
\newblock [,167(1975)]

\bibitem{Abbott:2016blz}
B.P. Abbott, et~al., Phys. Rev. Lett. \textbf{116}(6), 061102 (2016).
\newblock \doi{10.1103/PhysRevLett.116.061102}

\bibitem{Abbott:2016nmj}
B.P. Abbott, et~al., Phys. Rev. Lett. \textbf{116}(24), 241103 (2016).
\newblock \doi{10.1103/PhysRevLett.116.241103}

\bibitem{Abbott:2017vtc}
B.P. Abbott, et~al., Phys. Rev. Lett. \textbf{118}(22), 221101 (2017).
\newblock \doi{10.1103/PhysRevLett.118.221101, 10.1103/PhysRevLett.121.129901}.
\newblock [Erratum: Phys. Rev. Lett.121,no.12,129901(2018)]

\bibitem{Kokkotas:1999bd}
K.D. Kokkotas, B.G. Schmidt, Living Rev. Rel. \textbf{2}, 2 (1999).
\newblock \doi{10.12942/lrr-1999-2}

\bibitem{Berti:2009kk}
E.~Berti, V.~Cardoso, A.O. Starinets, Class. Quant. Grav. \textbf{26}, 163001
  (2009).
\newblock \doi{10.1088/0264-9381/26/16/163001}

\bibitem{Panotopoulos:2018can}
G.~Panotopoulos, Gen. Rel. Grav. \textbf{50}(6), 59 (2018).
\newblock \doi{10.1007/s10714-018-2381-5}

\bibitem{Panotopoulos:2017hns}
G.~Panotopoulos, A.~Rinc\'on, Int. J. Mod. Phys. \textbf{D27}(03), 1850034
  (2017).
\newblock \doi{10.1142/S0218271818500347}

\bibitem{Destounis:2018utr}
K.~Destounis, G.~Panotopoulos, A.~Rinc\'on, Eur. Phys. J. \textbf{C78}(2), 139
  (2018).
\newblock \doi{10.1140/epjc/s10052-018-5576-8}

\bibitem{Manfredi:2017xcv}
L.~Manfredi, J.~Mureika, J.~Moffat, Phys. Lett. \textbf{B779}, 492 (2018).
\newblock \doi{10.1016/j.physletb.2017.11.006}

\bibitem{Rincon:2018sgd}
A.~Rinc\'on, G.~Panotopoulos, Phys. Rev. \textbf{D97}(2), 024027 (2018).
\newblock \doi{10.1103/PhysRevD.97.024027}

\bibitem{Rincon:2018ktz}
A.~Rinc\'on, G.~Panotopoulos, Eur. Phys. J. \textbf{C78}(10), 858 (2018).
\newblock \doi{10.1140/epjc/s10052-018-6352-5}

\bibitem{Deser:1983tn}
S.~Deser, R.~Jackiw, G.~'t~Hooft, Annals Phys. \textbf{152}, 220 (1984).
\newblock \doi{10.1016/0003-4916(84)90085-X}

\bibitem{Deser:1988qn}
S.~Deser, R.~Jackiw, Commun. Math. Phys. \textbf{118}, 495 (1988).
\newblock \doi{10.1007/BF01466729}

\bibitem{tHooft:1988qqn}
G.~'t~Hooft, Commun. Math. Phys. \textbf{117}, 685 (1988).
\newblock \doi{10.1007/BF01218392}

\bibitem{Witten:1988hc}
E.~Witten, Nucl. Phys. \textbf{B311}, 46 (1988).
\newblock \doi{10.1016/0550-3213(88)90143-5}

\bibitem{Carlip:1995qv}
S.~Carlip, Class. Quant. Grav. \textbf{12}, 2853 (1995).
\newblock \doi{10.1088/0264-9381/12/12/005}

\bibitem{0264-9381-3-4-010}
J.D. Barrow, A.B. Burd, D.~Lancaster, Classical and Quantum Gravity
  \textbf{3}(4), 551 (1986).
\newblock \urlprefix\url{http://stacks.iop.org/0264-9381/3/i=4/a=010}

\bibitem{Banados:1992wn}
M.~Ba\~nados, C.~Teitelboim, J.~Zanelli, Phys. Rev. Lett. \textbf{69}, 1849
  (1992).
\newblock \doi{10.1103/PhysRevLett.69.1849}

\bibitem{Banados:1992gq}
M.~Ba\~nados, M.~Henneaux, C.~Teitelboim, J.~Zanelli, Phys. Rev. \textbf{D48},
  1506 (1993).
\newblock \doi{10.1103/PhysRevD.48.1506, 10.1103/PhysRevD.88.069902}.
\newblock [Erratum: Phys. Rev.D88,069902(2013)]

\bibitem{Cruz:1994ir}
N.~Cruz, C.~Mart\'inez, L.~Pe\~na, Class. Quant. Grav. \textbf{11}, 2731
  (1994).
\newblock \doi{10.1088/0264-9381/11/11/014}

\bibitem{Banados:1998gg}
M.~Ba\~nados, AIP Conf. Proc. \textbf{484}(1), 147 (1999).
\newblock \doi{10.1063/1.59661}

\bibitem{Cruz:2004vp}
N.~Cruz, S.~Lepe, Phys. Lett. \textbf{B593}, 235 (2004).
\newblock \doi{10.1016/j.physletb.2004.04.072}

\bibitem{Cardoso:2001hn}
V.~Cardoso, J.P.S. Lemos, Phys. Rev. \textbf{D63}, 124015 (2001).
\newblock \doi{10.1103/PhysRevD.63.124015}

\bibitem{Crisostomo:2004hj}
J.~Crisostomo, S.~Lepe, J.~Saavedra, Class. Quant. Grav. \textbf{21}, 2801
  (2004).
\newblock \doi{10.1088/0264-9381/21/12/002}

\bibitem{Cruz:1994ar}
N.~Cruz, J.~Zanelli, Class. Quant. Grav. \textbf{12}, 975 (1995).
\newblock \doi{10.1088/0264-9381/12/4/008}

\bibitem{garcia03}
A.A. Garc\'ia, C.~Campuzano, Phys. Rev. \textbf{D67}(6), 064014 (2003).
\newblock \doi{10.1103/PhysRevD.67.064014}

\bibitem{Cruz:2004tz}
N.~Cruz, M.~Olivares, J.R. Villanueva, Gen. Rel. Grav. \textbf{37}, 667 (2005).
\newblock \doi{10.1007/s10714-005-0054-7}

\bibitem{Cataldo:2006yk}
M.~Cataldo, N.~Cruz, Phys. Rev. \textbf{D73}, 104026 (2006).
\newblock \doi{10.1103/PhysRevD.73.104026}

\bibitem{Koch:2016uso}
B.~Koch, I.A. Reyes, A.~Rinc\'on, Class. Quant. Grav. \textbf{33}(22), 225010
  (2016).
\newblock \doi{10.1088/0264-9381/33/22/225010}

\bibitem{Rincon:2018dsq}
A.~Rinc\'on, E.~Contreras, P.~Bargue\~no, B.~Koch, G.~Panotopoulos, Eur. Phys.
  J. \textbf{C78}(8), 641 (2018).
\newblock \doi{10.1140/epjc/s10052-018-6106-4}

\bibitem{sdbtzsagnac}
A.~Rinc\'on, J.R. Villanueva,   (2019)

\bibitem{Rincon:2017ypd}
A.~Rinc\'on, B.~Koch, I.~Reyes, J. Phys. Conf. Ser. \textbf{831}(1), 012007
  (2017).
\newblock \doi{10.1088/1742-6596/831/1/012007}

\bibitem{Rincon:2017goj}
A.~Rinc\'on, E.~Contreras, P.~Bargue\~no, B.~Koch, G.~Panotopoulos,
  A.~Hern\'andez-Arboleda, Eur. Phys. J. \textbf{C77}(7), 494 (2017).
\newblock \doi{10.1140/epjc/s10052-017-5045-9}

\bibitem{Rincon:2017ayr}
A.~Rinc\'on, B.~Koch, J. Phys. Conf. Ser. \textbf{1043}(1), 012015 (2018).
\newblock \doi{10.1088/1742-6596/1043/1/012015}

\bibitem{Contreras:2017eza}
E.~Contreras, A.~Rinc\'on, B.~Koch, P.~Bargue\~no, Int. J. Mod. Phys.
  \textbf{D27}(03), 1850032 (2017).
\newblock \doi{10.1142/S0218271818500323}

\bibitem{Hernandez-Arboleda:2018qdo}
A.~Hern\'andez-Arboleda, A.~Rinc\'on, B.~Koch, E.~Contreras, P.~Bargue\~no,
  (2018)

\bibitem{Contreras:2018dhs}
E.~Contreras, A.~Rinc\'on, B.~Koch, P.~Bargue\~no, Eur. Phys. J.
  \textbf{C78}(3), 246 (2018).
\newblock \doi{10.1140/epjc/s10052-018-5709-0}

\bibitem{Rincon:2018lyd}
A.~Rinc\'on, B.~Koch, Eur. Phys. J. \textbf{C78}(12), 1022 (2018).
\newblock \doi{10.1140/epjc/s10052-018-6488-3}

\bibitem{Contreras:2018gct}
E.~Contreras, A.~Rinc\'on, J.M. Ram\'irez-Velasquez, Eur. Phys. J.
  \textbf{C79}(1), 53 (2019).
\newblock \doi{10.1140/epjc/s10052-019-6601-2}

\bibitem{Canales:2018tbn}
F.~Canales, B.~Koch, C.~Laporte, A.~Rinc\'on,   (2018)

\bibitem{Rincon:2019cix}
A.~Rinc\'on, E.~Contreras, P.~Bargue\~no, B.~Koch,   (2019)

\bibitem{Contreras:2018swc}
E.~Contreras, P.~Bargue\~no, Int. J. Mod. Phys. \textbf{D27}(09), 1850101
  (2018).
\newblock \doi{10.1142/S0218271818501018}

\bibitem{Contreras:2018gpl}
E.~Contreras, P.~Bargue\~no, Mod. Phys. Lett. \textbf{A33}(32), 1850184 (2018).
\newblock \doi{10.1142/S0217732318501845}

\bibitem{Stevenson:1981vj}
P.M. Stevenson, Phys. Rev. \textbf{D23}, 2916 (1981).
\newblock \doi{10.1103/PhysRevD.23.2916}

\bibitem{Reuter:2003ca}
M.~Reuter, H.~Weyer, Phys. Rev. \textbf{D69}, 104022 (2004).
\newblock \doi{10.1103/PhysRevD.69.104022}

\bibitem{Becker:2014qya}
D.~Becker, M.~Reuter, Annals Phys. \textbf{350}, 225 (2014).
\newblock \doi{10.1016/j.aop.2014.07.023}

\bibitem{Dietz:2015owa}
J.A. Dietz, T.R. Morris, JHEP \textbf{04}, 118 (2015).
\newblock \doi{10.1007/JHEP04(2015)118}

\bibitem{Labus:2016lkh}
P.~Labus, T.R. Morris, Z.H. Slade, Phys. Rev. \textbf{D94}(2), 024007 (2016).
\newblock \doi{10.1103/PhysRevD.94.024007}

\bibitem{Morris:2016spn}
T.R. Morris, JHEP \textbf{11}, 160 (2016).
\newblock \doi{10.1007/JHEP11(2016)160}

\bibitem{Ohta:2017dsq}
N.~Ohta, PTEP \textbf{2017}(3), 033E02 (2017).
\newblock \doi{10.1093/ptep/ptx020}

\bibitem{Chandrasekhar:579245}
S.~Chandrasekhar, \emph{{The mathematical theory of black holes}}.
\newblock Oxford classic texts in the physical sciences (Oxford Univ. Press,
  Oxford, 2002).
\newblock \urlprefix\url{https://cds.cern.ch/record/579245}

\bibitem{cov04}
N.~Cruz, M.~Olivares, J.R. Villanueva, Class. Quant. Grav. \textbf{22}, 1167
  (2005).
\newblock \doi{10.1088/0264-9381/22/6/016}

\bibitem{Villanueva:2018kem}
J.R. Villanueva, F.~Tapia, M.~Molina, M.~Olivares, Eur. Phys. J. \textbf{C78},
  10 (2018).
\newblock \doi{10.1140/epjc/s10052-018-6328-5}

\bibitem{Villanueva:2013gra}
J.R. Villanueva, Y.~V\'asquez, Eur. Phys. J. \textbf{C73}(10), 2587 (2013).
\newblock \doi{10.1140/epjc/s10052-013-2587-3}

\bibitem{Kuniyal:2015uta}
R.S. Kuniyal, R.~Uniyal, H.~Nandan, K.D. Purohit, Gen. Rel. Grav.
  \textbf{48}(4), 46 (2016).
\newblock \doi{10.1007/s10714-016-2041-6}

\end{thebibliography}

\end{document}